\documentstyle{aipproc} 

\input epsf.tex 

\newcommand{\mincir}{\raise -2.truept\hbox{\rlap{\hbox{$\sim$}}\raise5.truept 
\hbox{$?$}\ }} 
\newcommand{\gr}{\kern 2pt\hbox{}^\circ{\kern -2pt K}} 
\newcommand{\magcir}{\raise -2.truept\hbox{\rlap{\hbox{$\sim$}}\raise5.truept 
\hbox{$?$}\ }}

\newcommand{\be}{\begin{equation}} 
\newcommand{\ee}{\end{equation}} 
\newcommand{\bea}{\begin{eqnarray}} 
\newcommand{\eea}{\end{eqnarray}}

\newcommand{\etal}{{et al.}} 
\newcommand{\apj}{Astrophys. J.}

\newcommand{\mnras}{Mon. Not. Roy. Astron. Soc.} 
\newcommand{\aap}{Astron. and Astrophys.}

\def\be{\begin{equation}}
\def\ee{\end{equation}}

\def\bea{\begin{eqnarray}}
\def\eea{\end{eqnarray}}

\def\etal{{\it et al.}}

\begin{document}

\title{
Constraints on the neutrino mass \\ and the cosmological constant from \\
large scale structure observations
}

\author{B.~Novosyadlyj$^{*}$, R.~Durrer$^{\dagger}$ and S.~Apunevych$^{*}$}

\address{$^*$Astronomical Observatory of Ivan Franko National University of L'viv, Ukraine\\
$^{\dagger}$ Department de Physique Th\'eorique, Universit\'e de Gen\`eve, Switzerland\\
}

\maketitle 
 
\begin{abstract}
\noindent The observational data on the large scale structure (LSS) of the Universe 
are used to establish the upper limit on the neutrino content marginalized over
all other cosmological parameters within the class of adiabatic 
inflationary models. It is shown that the upper 2$\sigma$ limit
on the neutrino content can be expressed in the form 
$\Omega_{\nu}h^2/N_{\nu}^{0.64}\le0.042$ or, via the neutrino mass, 
$m_{\nu}\le4.0$eV. 
 
\end{abstract}  


\section*{Introduction} 
In the light of recent data on the matter content in the Universe
($0.25\le\Omega_m\le 0.5$) coming from cosmological observations
and lower limit on the neutrino contribution 
($\Omega_\nu\ge 0.001$) which follows from  the Super-Kamiokande 
experiment, the question whether cosmology still allows neutrino mass of
a few eV is of interest  for Cosmology as well as Particle Physics.
Relict massive neutrinos as collisionless particles with substantial
velocities suppress the power spectrum of matter density
perturbations  below of free-streaming scale, which depends 
mainly on the neutrino mass, the Hubble constant and time. Hence, the
comparison of the growth of mass 
fluctuations on these scales can be translated into an
 upper limit for the neutrino mass. From the Ly-$\alpha$ forest
in quasar spectra in together with a conservative implementation of
other cosmological constraints a limit of  $m_\nu\le5.5$eV
at 95\% C.L. has been reported~\cite{cr99}. The problem of this
determination is that the damping of small scale power due to
neutrinos can be imitated by variations of other cosmological
parameters like the spectral index $n_s$, the cosmological constant $\Lambda$,
 spatial curvature $\Omega_k$, the Hubble parameter $h$
($\equiv H_0/(100$km/s/Mpc$)$) and the baryon content $\Omega_b$.
Therefore, $\Omega_\nu$ (or $m_\nu$) must be determined simultaneously
with all these parameters using
a wide range scale of cosmological observations. Such investigations show
also that mixed dark matter models with cosmological constant ($\Lambda$MDM) 
can explain virtually all cosmological measurements \cite{nov00a}.  
The goal of this paper is to determine more carefully the upper limit on the neutrino
content and its rest mass in the framework of $\Lambda$MDM models.

We determine the parameters of the cosmological model which matches 
observational data on the large scale structure of the Universe best and,
marginalizing over all other parameters, we determine the upper limit
for the neutrino content and the neutrino mass. We restrict ourselves 
the sub-class of models without tensor mode and neglect early reionization.

\section*{Experimental data set and Method} 
 
Our approach is based on the comparison of the observational data on the 
structure of the Universe over a wide range of scales with theoretical 
predictions from the power spectrum of small (linear) density fluctuations.
The form of the spectrum strongly depends on the 
cosmological parameters $\Omega_m$, $\Omega_b$, $\Omega_{\nu}$, $N_{\nu}$,
$h$ and $n_s$. If the amplitude on a given scale is fixed by some
observational data then predictions for observations on all other scales
can be calculated and compared with 
 observational data. Minimizing quadratic differences 
between the theoretical and observational values divided by the observational 
errors, $\chi^2$, determines the best-fit values of the 
above mentioned cosmological parameters. We use the following
observational data set:

1. The location $\tilde\ell_p=197\pm6$ and amplitude $\tilde A_p=69\pm8\mu \rm K$
of the first acoustic  peak in the angular power spectrum of the CMB 
temperature fluctuations deduced from CMB map obtained in the Boomerang 
experiment \cite{ber00} (these data are sensitive to the amplitude and form of the
initial power spectrum in the range $\approx 200h^{-1}$Mpc);

2. The power spectrum of density fluctuations of Abell-ACO clusters ($\tilde P_{A+ACO}(k_j)$)
obtained from their space distribution by  \cite{ret97}
(scale range $10-100h^{-1}$Mpc);

3. The constraint for the amplitude of the fluctuation power spectrum on 
$\approx 10h^{-1}$Mpc scale derived from a recent optical determination of the 
mass function of nearby galaxy clusters \cite{gir98};
 
4. The constraint for the amplitude of the fluctuation power spectrum on 
$\approx 10h^{-1}$Mpc scale derived from the observed evolution of the galaxy 
cluster X-ray temperature distribution function between $z=0.05$ and $z=0.32$
\cite{via99}; 

5. The constraint for the amplitude of the fluctuation power spectrum on 
$\approx 10h^{-1}$Mpc scale derived from the existence of three very massive 
clusters of galaxies observed 
so far at $z>0.5$ \cite{bah98}; 

6. The constraint on the amplitude of the linear power spectrum of 
density fluctuations in our vicinity obtained from the study of  
bulk flows of galaxies in sphere of radius 
$50h^{-1}$Mpc  $\tilde V_{50}=(375\pm 85) {\rm km/s}$ 
\cite{kol97};

7. The constraint for the amplitude of the fluctuation power spectrum on 
$0.1-1h^{-1}$Mpc scale and $z=3$ derived from the Ly-$\alpha$ absorption lines seen in 
quasar spectra \cite{gn98};
 
8. The constraints for the amplitude and inclination of the initial power 
spectrum on $\approx 1h^{-1}$Mpc scale and $z=2.5$ scale derived 
from the Ly-$\alpha$ forest of quasar absorption lines~\cite{cr98};
 
9. Results from direct measurements of the Hubble constant 
$\tilde h=0.65\pm 0.10$ which is a compromise between 
by different groups; 

10. The nucleosynthesis constraint on the baryon density derived from a 
observational content of inter galactic deuterium,
$\widetilde{\Omega_bh^2} = 0.019\pm 0.0024$ (95\% C.L.) 
 \cite{bur99}.
 
One of the main ingredients for the solution of our search problem 
is a reasonably fast and accurate determination of the  linear 
transfer function for dark matter clustering 
which depends on the cosmological parameters. We use accurate analytical 
approximations of the MDM transfer function $T(k;z)$ depending on 
the parameters $\Omega_m$, $\Omega_b$, $\Omega_{\nu}$, $N_{\nu}$ and 
$h$ given in Ref.~\cite{eh3}. The 
linear power spectrum of matter density fluctuations is
$P(k;z)=A_sk^{n_s}T^2(k;z)D_1^2(z)/D_1^2(0)$, 
where $A_s$ is the normalization constant for scalar perturbations and 
$D_1(z)$ is the linear growth factor. 
We normalize the spectra to the 4-year COBE data 
which determine the amplitude of density perturbation at horizon 
scale, $\delta_h$ \cite{bun97}. Therefore, each model will match
the COBE data by construction. The normalization constant $A_s$ is 
then given by $A_s=2\pi^{2}\delta_{h}^{2}(3000{\rm Mpc}/h)^{3+n_s}$. 
 
The Abell-ACO power spectrum is related to the matter power 
spectrum at $z=0$ by the cluster biasing parameter $b_{cl}$:
$P_{A+ACO}(k)=b_{cl}^{2}P(k;0)$. We assume scale-independent linear
bias as free parameter the best-fit value of which will be determined
together with the other cosmological parameters. 
 
The dependence of the position and amplitude of the first acoustic 
peak in the CMB power spectrum on cosmological 
parameters $n_s$, $h$, $\Omega_b$, $\Omega_{cdm}$ and 
$\Omega_{\Lambda}$ can be determined using an analytical approximation 
given in \cite{efs99} which has been  extended to  models with non-zero 
curvature  ($\Omega_k\equiv1-\Omega_m-\Omega_{\Lambda}\ne0$) in 
\cite{dur00}. Its accuracy in the parameter range 
considered is better then 5\%.

The theoretical values of the other experimental constraints are
calculated as described in \cite{nov00a}. There
one can also find  some tests of our method.

\section*{Results and Discussion}

The determination of the parameters  
$\Omega_m$, $\Omega_{\Lambda}$, $\Omega_{\nu}$, $N_{\nu}$, $\Omega_b$,  
$h$, $n_s$ and $b_{cl}$ by the Levenberg-Marquardt $\chi^2$ 
minimization method is realized as follows: we vary the 
set of parameters  
$\Omega_m$, $\Omega_{\Lambda}$, $\Omega_{\nu}$, $\Omega_b$,  
$h$, $n_s$ and $b_{cl}$ with fixed $N_{\nu}$ (= number of massive
neutrino species) and find the minimum of 
$\chi^2$. Since $N_{\nu}$ is discrete we repeat this 
procedure for $N_{\nu}$=1, 2, and 3.  The lowest of the 
three minima is the minimum of $\chi^2$ for the 
complete set of free parameters. We have seven continous free parameters. 
The formal number of data points is 24 but, as it was shown
in \cite{nov00a}, the 13 points of the cluster power spectrum  can be
described by just 3 degrees of freedom, so that the  
maximal number of truly independent measurements is 14. Therefore,  
the number of degrees of 
freedom for our search procedure is $N_F= N_{\rm exp}-N_{\rm par}= 7$.  
The  
model with one sort of massive neutrinos provides the best fit to  
the data, $\chi^2_{min}=5.9$. Its parameters are following
$\Omega_m=0.37^{+0.25}_{-0.15}$,      
$\Omega_{\Lambda}=0.69^{+0.15}_{-0.20}$,      
$\Omega_{\nu}=0.03^{+0.07}_{-0.03}$,    $N_{\nu}=1$,  
$\Omega_b=0.037^{+0.033}_{-0.018}$,   
$n_s=1.02^{+0.09}_{-0.10}$,      
$h=0.71^{+0.22}_{-0.19}$,      
$b_{cl}=2.4^{+0.7}_{-0.7}$. The errors  
 are obtained by maximizing the (Gaussian) 68\%
confidence contours over all other parameters.
  
However,  there is  
only a marginal difference in $\chi^2_{min}$ for $N_\nu =1,2,3$.    
With the given accuracy of the data we cannot conclude  
whether  massive neutrinos are present  
at all, and if yes what number of degrees of freedom is favored.   
We summarize, that the  
considered observational data on LSS of the Universe can be  
explained by a $\Lambda$MDM inflationary model with approximately 
a scale invariant spectrum of scalar perturbations and small positive 
curvature, $\Omega_{k}=-0.06$. It is interesting to note that
the values of the fundamental cosmological parameters 
$\Omega_m$, $\Omega_{\Lambda}$ and  
$\Omega_k$  determined by these observations of large 
scale structure match the SNIa test 
$\Omega_m-0.75\Omega_{\Lambda}=-0.25\pm0.125~$ \cite{per98}
very well. Models with vanishing $\Lambda$ are outside of marginalized 
$3 \sigma$ contour of the best-fit model with $N_\nu=1$ determined by the LSS 
observational characteristics used here even without the SNIa 
constraint (Fig. 1a). For models with $N_\nu=2$ and $N_\nu=3$ contours are
a bit wider, so models with vanishing $\Lambda$ are outside the marginalized 
$2 \sigma$ contour for arbitrary $N_\nu$.
 
Results change only 
slightly if instead of the Boomerang data we use Boomerang+MAXIMA-1. 
Hence, we  can conclude that the LSS observational  
characteristics together with the Boomerang (+MAXIMA-1) data on the 
first acoustic peak already rule out  
zero-$\Lambda$ models at more than 95\% C.L.  and actually demand a 
cosmological constant in the same bulk part as direct measurements. We 
consider this a non-trivial consistency check! 
 
The neutrino matter density $\Omega_{\nu}=0.03$ 
corresponds to a neutrino mass of $m_{\nu}=94\Omega_{\nu}h^2\approx1.4$ eV 
but is compatible with 0 within $1\sigma$. 
A $\Lambda$CDM model ($\Omega_{\nu}=0$) is within the  1$\sigma$
contour of the best-fit $\Lambda$MDM model.  
 
To derive an upper limit for neutrino content and its rest mass,
we first determine the marginalized
1$\sigma$, 2$\sigma$ and 3$\sigma$ upper limits for $\Omega_{\nu}$ for
different values of $N_{\nu}$. Using the best-fit value for $h$ at
given  $\Omega_{\nu}$, we  then obtain the a corresponding upper
limit for the neutrino mass,  
$m_{\nu}=94\Omega_{\nu}h^2/N_{\nu}$. The results are presented in
Table 1 and Figure 1.  For more species of massive neutrino the upper limit for
$\Omega_{\nu}$ is somewhat higher but $m_{\nu}$ is still lower for
each C.L. The  upper limit for 
$\Omega_{\nu}$ raises with the confidence level as expected. But the
upper limit for the mass grows only very little due to the reduction
of the best-fit value for $h$. The upper limit for 
the combination  $\Omega_{\nu}h^2/N_{\nu}^{0.64}$ is
approximately constant for all numbers of species and confidence levels. 
The observational data
set used here establishes an upper limit for the massive neutrino
content of the universe which can be expressed in the form
$\Omega_{\nu}h^2/N_{\nu}^{0.64}\le0.042$  at 2$\sigma$ confidence 
level.
The corresponding upper limit on the neutrino mass $m_{\nu}\le4$eV 
is lower than the value obtained by \cite{cr99}.

\begin{table} 
\caption{Upper limits for the neutrino content and mass (in eV) at 
different confidence levels.  
\label{nu} }  
\def\onerule{\noalign{\medskip\hrule\medskip}} 
\medskip 
\begin{tabular}{|c|cc|cc|cc|} 
\hline 
&\multicolumn{2}{c|}{1$\sigma$ C.L.} 
&\multicolumn{2}{c|}{2$\sigma$ C.L.} 
&\multicolumn{2}{c|}{3$\sigma$ C.L.}\\ 
$N_{\nu}$ &$\Omega_{\nu}$ &$m_{\nu}$ &$\Omega_{\nu}$ &$m_{\nu}$ &$\Omega_{\nu}$ &$m_{\nu}$   \\ [4pt] 
\hline 
1&0.10&3.65&0.13&3.96&0.18&4.04\\ 
2&0.15&2.79&0.21&3.06&0.29&3.35\\ 
3&0.20&2.40&0.27&2.67&0.35&2.78\\ 
\hline 
\end{tabular} 
\end{table}

\begin{figure}[tp]
\epsfxsize=14.8truecm
\epsfbox{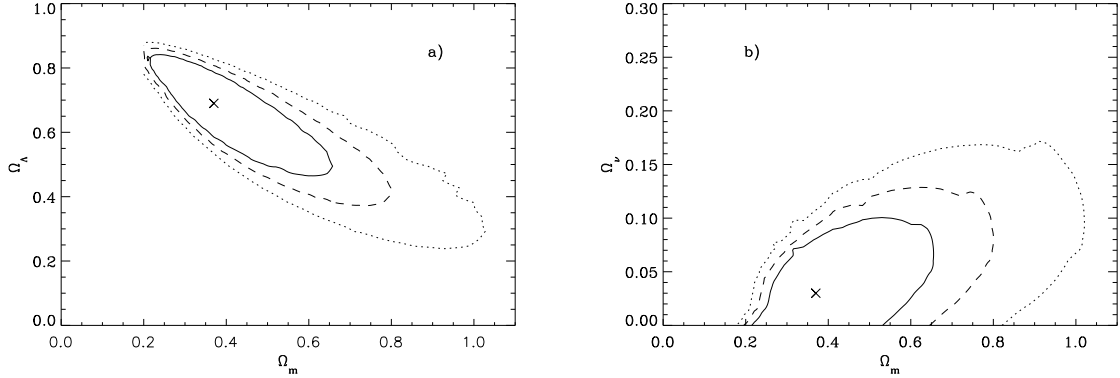}
\caption{Likelihood contours (solid line - 68.3\%, dashed - 95.4\%, dotted
- 99.73\%) of $\Lambda$MDM with $N_{\nu}=1$ in the
$\Omega_m-\Omega_\Lambda$ (a) and $\Omega_m-\Omega_{\nu}$ (b) planes
marginalized over all other parameters.}
\end{figure}


{\it Acknowledgments:}

B. Novosyadlyj is grateful to Geneva University for hospitality and to
the Swiss NSF for a grant for the participation in CAPP2000.

\end{document}